\documentclass[aps,pra,floatfix,reprint,
superscriptaddress,
longbibliography]{revtex4-1}
\usepackage{blindtext}
\usepackage{braket}
\usepackage{graphicx} 
\usepackage[T1]{fontenc}
\usepackage{amsmath,amsthm,amsfonts}
\usepackage{bm}
\usepackage{varioref}
\usepackage{soul}
\usepackage{dsfont}
\usepackage[colorlinks=true,linkcolor=blue,citecolor=red,filecolor=blue]{hyperref}
\usepackage{cleveref}
\usepackage{tikz}
\usepackage{xcolor}
\newcommand{\mycomment}[1]{}
\usetikzlibrary{arrows.meta, positioning, calc, fit}
\usepackage{comment}

\newtheorem{proposition}{Proposition}
\newtheorem{corollary}{Corollary}
\newcommand{\e}{\mathrm{e}}

\newcommand{\Tr}{\mathrm{Tr}}
\renewcommand{\Re}{\mathrm{Re}}
\renewcommand{\Im}{\mathrm{Im}}

\renewcommand{\thesection}{\Roman{section}}

\begin{document}
\title{Geometric bounds on multiparameter Heisenberg scaling in optical metrology with limited squeezed resources}
\author{Atmadev Rai}
\affiliation{School of Mathematics and Physics, University of Portsmouth, Portsmouth PO1 3QL, UK}
\affiliation{Quantum Science and Technology Hub, University of Portsmouth, Portsmouth PO1 3QL, United Kingdom}
\author{Paolo Facchi}
\affiliation{Dipartimento di Fisica, Universit\`{a} di Bari, I-70126 Bari, Italy}
\affiliation{INFN, Sezione di Bari, I-70126 Bari, Italy}
\author{Vincenzo Tamma}
\email[]{vincenzo.tamma@port.ac.uk}
\affiliation{School of Mathematics and Physics, University of Portsmouth, Portsmouth PO1 3QL, UK}
\affiliation{Quantum Science and Technology Hub, University of Portsmouth, Portsmouth PO1 3QL, United Kingdom}
\affiliation{Institute of Cosmology and Gravitation, University of Portsmouth, Portsmouth PO1 3FX, UK}
\begin{abstract}
The simultaneous estimation of multiple parameters is a central task in quantum metrology, distributed sensing, and the calibration of large photonic interferometers. A fundamental question is how many independent parameter combinations can inherit Heisenberg scaling from a given number of squeezed probes in a multimode Gaussian network. Here, we answer this question for arbitrary passive linear optical networks. For a $p$-parameter, $M$-channel interferometer probed by $k$ single-mode squeezed states and at least one coherent state in the remaining channels, we show that the rank of the Heisenberg-scaling coefficient of the quantum Fisher information matrix is bounded by $n_{\rm HS}\le \min\{p,k(k+3)/2\}$, which corresponds to the maximum number of independent combinations of parameters that can be estimated with Heisenberg-scaling sensitivity. The bound separates into two geometrically distinct contributions. The covariance contribution of the quantum Fisher information, which describes squeezing-enhanced fluctuations, provides at most $k(k+1)/2$ parameter combinations estimable at Heisenberg-scaling sensitivity, while the first-moment contribution provides at most $k$ additional independent parameter combinations with Heisenberg-scaling sensitivity. We identify the conditions for saturating these bounds and construct a passive family of interferometers that saturates these bounds.
\end{abstract}

\maketitle
\twocolumngrid
\section{Introduction}
Large optical networks may encode many optical phases, with the number of unknown parameters growing rapidly with network size. In practice, however, the number of nonclassical probes is limited. An optical network can redistribute the input light across multiple output ports, but it cannot allow an unlimited number of independent parameters to gain quantum metrological advantage such as Heisenberg-scaling precision $O(1/N^2)$ with the total average input photon number $N$~\cite{giovannetti2004quantum,PhysRevLett.71.1355,PhysRevLett.96.010401,PhysRevResearch.1.032024,giovannetti2011advances}. This sets a fundamental limit on the number of independent parameter combinations that can achieve Heisenberg scaling from a finite number of nonclassical probes such as squeezed states of light~\cite{Andersen_2016,Loudon01061987, PhysRevD.23.1693,PhysRevD.30.2548,PhysRevLett.68.3020}.

This issue is central to multiparameter quantum metrology, where several unknown parameters are encoded simultaneously in multimode photonic networks~\cite{szczykulska2016multi, ALBARELLI2020126311, PhysRevLett.111.070403, PhysRevLett.120.080501,PhysRevA.94.042342,y9wq-hhg2,PhysRevA.104.062603,PhysRevA.111.062408,rai2025simultaneous, Gramegna_2021}. The transition from single-parameter estimation to multiparameter settings is driven by the importance of its applications such as distributed sensing, quantum imaging, and the characterization of large optical interferometric networks~\cite{PhysRevLett.120.080501, szczykulska2016multi, PhysRevLett.121.043604}. Multiphase interferometric schemes have shown that simultaneous estimation can outperform separate strategies and that squeezed or entangled resources can enhance the estimation of parameters across many channels~\cite{PhysRevA.104.062603}. These results establish how well a chosen set of parameters can be estimated in a given setting. They do not, however, determine how many independent parameter combinations can, in principle, obtain enhanced precision of Heisenberg scaling in a generic passive multimode network with only a limited number of nonclassical resources. Given a unitary with $p$ unknown parameters, we show what the maximum number of parameters or independent parameter combinations one can estimate with Heisenberg scaling for a given number of single-mode squeezed states (SMSS) at the input is. As a direct consequence, we also find a minimum number of SMSS needed to simultaneously estimate all $p$ parameters in an optical network at Heisenberg-scaling precision. The main idea of this work is therefore not the optimization over states or measurements, but a parameterization-independent bound on the dimension of the leading Heisenberg-scaling subspace of the Fisher matrix. 

Continuous-variable optical networks, in particular multimode linear interferometers with experimentally feasible and tunable optical elements, provide a natural platform to explore such multiparameter quantum metrology~\cite{RevModPhys.77.513, RevModPhys.84.621}. In Gaussian metrology, the Fisher information can be expressed directly in terms of the evolution of the first and second moments, which makes the phase-space structure of the problem explicit~\cite{monras2013phase}.

\begin{figure*}
    \centering
    \includegraphics[width=\textwidth]{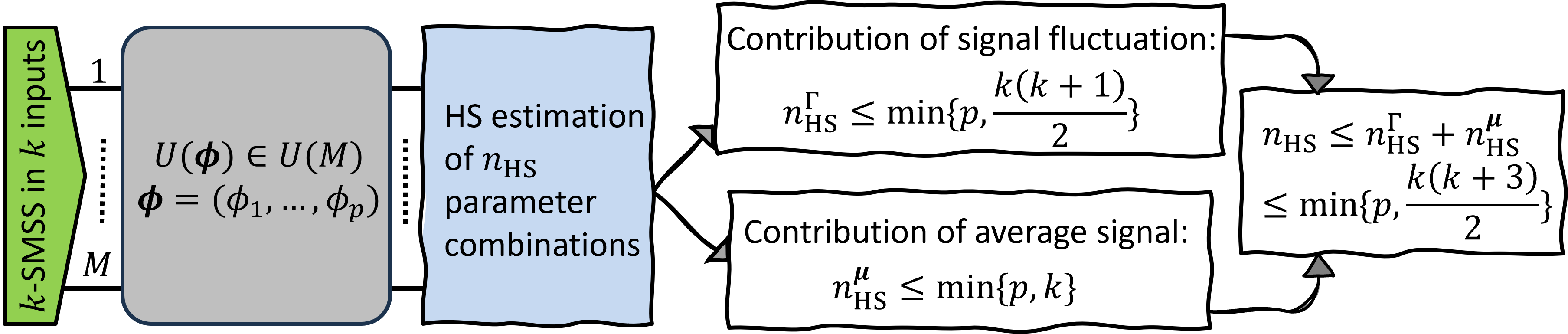}
    \caption{Schematic of a generic $M$-channel optical sensor with $p$ arbitrarily encoded parameters and $k$ injected single-mode squeezed states (SMSSs). Due to the Gaussian nature of probe and passive optical network, the quantum Fisher information matrix separates into a covariance contribution $\mathcal F^{\Gamma}$ in Eq.~\eqref{eq:F_cov_main}, associated with parameter-dependent fluctuations, and a first-moment (mean) contribution $\mathcal F^{\bm\mu}$ in Eq.~\eqref{eq:F_mean_main}, associated with parameter-dependent displacements of the Gaussian center. The figure summarizes the resulting bounds on Heisenberg-scaling (HS) precision. The covariance contribution of QFIM can obtain at most $n_{\rm HS}^\Gamma$ independent parameter combinations with HS sensitivity as given in Eq.~\eqref{eq:main_cov_bound}. The first-moment contribution can achieve HS sensitivity for at most $n_{\rm HS}^{\bm \mu}$ independent parameter combinations given in Eq.~\eqref{eq:main_mean_bound}. Taken together, the full QFIM can therefore exhibit HS scaling for at most $n_{\rm HS}$ independent parameter combinations in Eq.~\eqref{eq:main_full_bound}.}
    \label{fig:double_column}
\end{figure*}

In this article, we answer this question for passive multimode Gaussian metrology by considering an arbitrary $p$-parameter passive linear network $U(\bm\phi)\in \rm U(M)$ in Fig.~\ref{fig:double_column} injected with a multimode Gaussian probe containing $k$-SMSS and vacuum or coherent states in the remaining channels. We show that the leading Heisenberg-scaling precision in the total number of input photons is obtained using the quantum Fisher information matrix (QFIM)~\cite{helstrom1969quantum, PhysRevLett.72.3439, demkowicz2020multi}, limited by the active input squeezed subspace. More precisely, we prove that the covariance contribution of the QFIM can obtain at most $k(k+1)/2$ independent Heisenberg-scaling-enabled parameter combinations, and the first-moment contribution gives at most $k$ parameter combinations; therefore, the full QFIM allows at most $k(k+3)/2$ independent parameter combinations withe Heisenberg-scaling sensitivity, which cleanly separates the roles of signal fluctuations and signal mean-field contributions of the Fisher matrix in continuous-variable quantum metrology. 
Our result is not an optimization over a particular measurement or over a chosen set of phases. Rather, it is a parameterization-independent bound on the dimension of the
subspace of parameter combinations whose Fisher information eigenvalues scale as $N^2$. We also identify saturation conditions for these geometric bounds and construct finite passive interferometers that reach the bound for the QFIM.

\section{Multiparameter Gaussian model}
\label{sec:FIM_main}
An $M$-mode Gaussian state is fully described by its first and second moments, namely the mean vector $\bm{\mu}$ and the covariance matrix $\Gamma$. In phase space, its Wigner function reads
\begin{equation}
  W(\bm z)
  =
  \frac{1}{(2\pi)^M\sqrt{\det\Gamma}}\,
  \exp\!\left[-\frac12
    \big(\bm z-\bm \mu\big)^{T}
    \Gamma^{-1}
    \big(\bm z-\bm \mu\big)
  \right],
  \label{eq:Gaussian_pdf_main}
\end{equation}
where $\bm z\in\mathbb{R}^{2M}$ denotes the phase-space coordinate vector. All parameter dependence in a Gaussian sensing protocol is therefore carried by the evolution of $\bm \mu$ and $\Gamma$~\cite{PhysRev.40.749}. We consider an arbitrary passive linear optical network $U(\bm\phi)\in \rm U(M)$, depending on a vector of real parameters $\bm\phi=(\phi_1,\dots,\phi_p)$. In phase space, $U(\bm\phi)$ is represented by a real symplectic matrix 
\begin{equation}
    R_U=\begin{pmatrix}
        \Re [U(\bm\phi)] & -\Im [U(\bm\phi)]\\
\Im [U(\bm\phi)] & \Re [U(\bm\phi)]
    \end{pmatrix},
    \label{eq:RU}
\end{equation}
therefore the input mean vector $\bm\mu_{\rm in}$ and covariance matrix $\Gamma_{\rm in}$ transform as $\bm\mu(\bm\phi) = R_U\,\bm\mu_{\rm in}$, and $\Gamma(\bm\phi) = R_U\,\Gamma_{\rm in}\,R^{\mathsf T}_U$ respectively. The parametric dependence of the interferometer network $U(\bm\phi)$ is given by the generators $\tau_i = R_U^{\mathsf T}(\partial_i R_U)$, and can be written in the standard decomposition of the symplectic algebra as
\begin{equation}
\tau_i =
\begin{pmatrix}
A_i & -B_i\\
B_i & A_i
\end{pmatrix},\qquad i=1,\dots,p,
\label{Eq:Generator_tau}
\end{equation}
where the antisymmetric block $A_i^\mathsf T=-A_i$ mixes modes within the same quadratures, while the symmetric block $B_i^\mathsf T=B_i$ mixes $q$- and $p$-quadratures. 

We assume that $k$ of the input modes are prepared in single-mode squeezed vacua and that the remaining $M-k$ modes are in coherent states or vacuum. For simplicity, we assume equal squeezing parameter $r$ in each mode, with $N_s=k\sinh^2r$ being the total squeezed number of photons. The results we present here are unchanged if unequal squeezing is considered, provided they scale with the same asymptotic resource $\sinh^2{r_a}\sim\rho_a N_s$ where $\rho_a>0$ (See Supplemental Material (SM)~\ref{Appendix_D1}).  In a suitable mode ordering, the input covariance can be written as
\begin{equation}
\Gamma_{\rm in}
=
\frac12\begin{pmatrix}
d & 0\\
0 & d^{-1}
\end{pmatrix},
\qquad
d = sP + Q,
\label{eq:Squeezed_input}
\end{equation}
where $P=\operatorname{diag}(1,\dots,1,0,\dots,0)$ projects onto the $k$ squeezed modes, $Q=\mathbb I_M-P$ projects onto the remaining $M-k$ modes, and $s = e^{2r}$. The input coherent state is described by an input mean vector of the form  $\bm\mu_{\rm in}=\sqrt{2N_c}\left(\bm q,\;\bm p\right)^\mathsf T$, where the real vectors $\bm q$ and $\bm p$ specify how the coherent amplitude is distributed over the modes and $N_c$ is the total mean photon number in the input coherent states.

For the pure Gaussian model, the quantum Fisher information matrix (QFIM) separates into a covariance contribution and a first-moment contribution $\mathcal F=
\mathcal F^{\Gamma}+\mathcal F^{\bm\mu}$, with matrix elements in terms of the generator $\tau$ in Eq.~\eqref{Eq:Generator_tau}
\begin{align}
\mathcal F^{\Gamma}_{ij}
&=
\frac14
\Tr\!\left[
(\tau_i-\Gamma_{\rm in}\tau_i\,\Gamma_{\rm in}^{-1})
(\tau_j-\Gamma_{\rm in}\tau_j\,\Gamma_{\rm in}^{-1})
\right],
\label{eq:F_cov_main}\\
\mathcal F^{\bm \mu}_{ij}
&=
4N_c
\left(
\xi_i^\mathsf{T} d^{-1}\xi_j
+
\eta_i^\mathsf{T} d\,\eta_j
\right),
\label{eq:F_mean_main}
\end{align}
where $\xi_i = A_i\bm q - B_i\bm p$, and
$\eta_i = B_i\bm q + A_i\bm p$. This decomposition has a clear physical meaning. The term $\mathcal F^{\Gamma}$ quantifies how the parameter variations transform the covariance matrix, i.e., the orientation of the noise ellipsoid in phase space and therefore captures information that is encoded in parameter-dependent fluctuations, which can be amplified by squeezing. The second term $\mathcal F^{\bm\mu}$ quantifies how parameters displace the Gaussian center and captures information carried by the amplified displacement of the mean field. Both terms can contribute to Heisenberg scaling of measurement precision, but they use different degrees of freedom in the squeezed input subspace. See sections~\ref{app:HS_scaling} and~\ref{app:HS_scaling_mean} of SM~\cite{Supp} for the detailed derivation of Eqs.~\eqref{eq:F_cov_main} and~\eqref{eq:F_mean_main}.

\section{Multiparameter Heisenberg scaling}
Let $N=N_s + N_c$ denote the total average input photon number, with $N_s$ being the photon number associated with the squeezed probe, and $N_c$ the photon number associated with the input coherent state. First, we analyze the covariance contribution of the QFIM in Eq.~\eqref{eq:F_cov_main}. Using Eq.~\eqref{eq:Squeezed_input} and expanding $\mathcal F^{\Gamma}$ in asymptotic powers of $s$, one finds that the leading $s^2\sim N_s^2$ term depends only on the projection of the symmetric generator block $B_i$ in~\eqref{Eq:Generator_tau} onto the squeezed input modes. All other terms are at most linear in $s$. This yields the asymptotic form (see section~\ref{app:HS_scaling} of SM)
\begin{equation}
\mathcal F^{\Gamma}=
N_s^2\,\mathcal F^{\Gamma}_{\rm HS}
+
O(N_s),
\label{eq:asym_cov_main}
\end{equation}
where the coefficient matrix $\mathcal F_{\rm HS}^{\Gamma}$ is independent of the average photon number associated with the Heisenberg scaling (HS) term. Up to an overall positive prefactor, 
this coefficient matrix is the Gram matrix
\begin{equation}
\mathcal F_{\rm HS}^{\Gamma} \propto \left[\Tr(C_iC_j)\right]_{i,j=1}^p,\qquad
    C_i=PB_iP,
    \label{eq:covariance_Gram}
\end{equation}
where $C_i$ are real symmetric $k\times k$ matrices in $\mathrm{Sym}(k,\mathbb R)$. Therefore its rank is the dimension of \(\operatorname{span}\{C_i\}\), which cannot exceed \(\dim\mathrm{Sym}(k,\mathbb R)=k(k+1)/2\) or the number of parameters~$p$.

We define the number of Heisenberg-scaling directions as the number of independent
parameter combinations whose Fisher-information eigenvalues scale as $N^2$. Equivalently,
if
$\mathcal F=N^2\mathcal F_{\rm HS}+O(N),$
then this number is $n_{\rm HS}=\operatorname{rank}\mathcal F_{\rm HS}$. Thus, at
leading order, multiparameter Heisenberg scaling is reduced to a finite-dimensional rank
problem on the active squeezed input subspace.
\begin{proposition}
\label{prop:cov_bound_main}
Let $U(\bm\phi)\in \rm U(M)$ be an arbitrary $p$-parameter passive linear optical network. For an input with $k$ single-mode squeezed states and coherent or vacuum states in the remaining modes,  the number $n_{\mathrm{HS}}^{\Gamma}$ of Heisenberg-scaling parameter combinations carried by the covariance contribution $\mathcal{F}^{\Gamma}$ of the QFIM  satisfies
\begin{equation}
n_{\mathrm{HS}}^{\Gamma}=\operatorname{rank}\!\big(\mathcal F^{\Gamma}_{\rm HS}\big)
\le
\min\!\left\{p,\frac{k(k+1)}{2}\right\}.
\label{eq:main_cov_bound}
\end{equation}
\end{proposition}
The covariance contribution $\mathcal F^\Gamma$ 
is fluctuation-driven and does not require any coherent
displacement. Its Heisenberg-scaling directions are determined by the compressed symmetric
blocks $C_i=PB_iP$, which live in $\mathrm{Sym}(k,\mathbb R)$. This space has dimension
$k(k+1)/2$, giving the bound. See section~\ref{App:Proposition_1} of SM for the proof. 

Next, we analyze the first-moment contribution of the QFIM $\mathcal F^{\bm \mu}$ in Eq.~\eqref{eq:F_mean_main}. Using Eqs.~\eqref{Eq:Generator_tau} and~\eqref{eq:Squeezed_input}, the first-moment contribution takes the asymptotic form 
\begin{equation}
    \mathcal F^{\bm \mu}=
N_sN_c\,\mathcal F^{\bm \mu}_{\rm HS}
+
O(N),
\label{eq:asym_mean_main}
\end{equation}
where the coefficient matrix $\mathcal F^{\bm \mu}_{\rm HS}$ is independent of the average photon number. Up to an overall positive prefactor, it is a Gram matrix 
\begin{equation}
\mathcal F^{\bm \mu}_{\rm HS}\propto\left[\bm g_i^\mathsf T\bm g_j\right]_{i,j=1}^p,
\qquad
    \bm g_i=P(B_i\bm q+A_i \bm p).
    \label{eq:mean_Gram_vector}
\end{equation}
Thus, the first-moment contribution is fully determined by the parameter-induced displacement projected onto the squeezed input subspace determined by the family of projected displacement vectors $\{\bm g_i\}\subset\mathbb R^k$. It contributes at order $N^2$ only when both the average number of squeezed photons $N_s$ and coherent photons $N_c$ scale linearly with the total photon number. See the derivation in section~\ref{app:HS_scaling_mean} of SM~\cite{Supp}. This gives the following rank bound.
\begin{proposition}
\label{prop:mean_bound_main}
Under the same assumptions of Proposition~\ref{prop:cov_bound_main}, and assuming that the coherent photon number $N_c$ scales linearly with the squeezed photon number $N_s$, the number $n^{\bm{\mu}}_{\mathrm{HS}}$ of Heisenberg-scaling independent parameter combinations  carried by the first-moment contribution $\mathcal{F}^{\bm{\mu}}$ satisfies
\begin{equation}
n_{\mathrm{HS}}^{\bm{\mu}}=\operatorname{rank}\!\big(\mathcal F^{\bm \mu}_{\rm HS}\big)
\le
\min\{p,k\}.
\label{eq:main_mean_bound}
\end{equation}
\end{proposition}
The coefficient matrix $\mathcal F^{\bm \mu}_{\rm HS}$ can contribute at Heisenberg scaling order only through a squeezed-amplified coherent displacement state within the same $k$-dimensional squeezing subspace. A non-zero input displacement is necessary for the mean contribution of the QFIM to give Heisenberg scaling. However, this bound in general does not require any fixed location of the input coherent displacement. The proof is given in section~\ref{app:HS_scaling_mean} of SM. 

Combining the two contributions of the QFIM in Eqs.~\eqref{eq:asym_cov_main} and~\eqref{eq:asym_mean_main}, and for the resource split $N_s=\beta N$ and $N_c=(1-\beta) N$, where $\beta\in (0,1)$, the full QFIM can be written as
\begin{equation}
    \mathcal{F} = N^2 \mathcal{F}_{\mathrm{HS}} + O(N),
    \label{eq:totalFIM}
\end{equation}
where $\mathcal F_{\rm HS}=\beta^2\mathcal F_{\rm HS}^{\Gamma}+\beta(1-\beta)\mathcal F_{\rm HS}^{\bm \mu}$. Therefore, the number of independent parameter combinations that retain Heisenberg scaling is $n_{\rm HS}=\operatorname{rank}(\mathcal{F}_{\mathrm{HS}})$. We now state the full resulting bound.
\begin{corollary}
\label{cor:full_bound_main}
For the full QFIM, the total number of independent parameter combinations with Heisenberg-scaling sensitivity is bounded by the combined ranks of the covariance and first-moment contributions, $n_{\rm HS}\le n_{\rm HS}^{\Gamma}+n_{\rm HS}^{\bm \mu}$. Hence,
\begin{equation}
n_{\mathrm{HS}}
\leq\min\!\left\{p,\frac{k(k+3)}{2}\right\}.
\label{eq:main_full_bound}
\end{equation}
\end{corollary}
The first inequality keeps track of how many first-moment directions are actually present;
the second gives the universal maximum. Therefore, the full bound is saturated only when the covariance contribution spans $\mathrm{Sym}(k,\mathbb R)$ and the first-moment contribution adds $k$ independent directions not already contained in $n^{\Gamma}_{\mathrm{HS}}$. This follows directly from Eqs.~\eqref{eq:main_cov_bound} and~\eqref{eq:main_mean_bound}, as the full Fisher matrix at the Heisenberg scaling in Eq.~\eqref{eq:totalFIM} is the sum of the covariance and first-moment coefficient matrices, so its rank cannot exceed the sum of their ranks, nor the number of parameters $p$. Combining Propositions~\ref{prop:cov_bound_main} and~\ref{prop:mean_bound_main} gives Eq.~\eqref{eq:main_full_bound}. The number of Heisenberg-scaling parameter combinations is therefore fixed by the dimensions of these two spaces, not by the number of modes or encoded parameters alone. Thus, with \(k\) SMSS, one cannot estimate more than $k(k+3)/2$ independent combinations of Heisenberg-scaling enabled parameters, regardless of how the parameters are encoded in $U(\bm\phi)$ (see section~\ref{App:proof_of_Corollary_1} of SM for a detailed proof).

A direct consequence is a lower bound on the number of SMSS required to achieve the Heisenberg-scaling precision for all $p$ parameters in a network.
\begin{corollary}
\label{cor:min_squeezers_full_hs}
Under the assumptions of Corollary~\ref{cor:full_bound_main}, estimating all $p$ parameters in a unitary $U(\bm\phi)$ with Heisenberg-scaling precision, i.e., 
$\operatorname{rank}\!\big(\mathcal F_{\rm HS}\big)=p$,
is possible only if
\begin{equation}
k\geq k_{\min}(p) =
\left\lceil
\frac{\sqrt{8p+9}-3}{2}
\right\rceil.
\label{eq:kmin_full_rank}
\end{equation}
\end{corollary}
Thus, if $k<k_{\min}(p)$, the coefficient matrix $\mathcal F_{\rm HS}$ is necessarily rank deficient, independently of the choice of unitary network $\rm U(M)$.
This follows directly from Corollary~\ref{cor:full_bound_main} as the full Heisenberg-scaling rank requires $k(k+3)/2\ge p$, and solving this quadratic inequality for the non-negative integer $k$ gives Eq.~\eqref{eq:kmin_full_rank}. For large $p$, the minimum number of SMSSs required for all $p$ parameters to achieve Heisenberg scaling grows only as $\sqrt{p}$. (see section~\ref{App:proof_of_Corollary_1} of SM~\cite{Supp}).

\section{Physical interpretation of the saturation conditions}
\label{sec:Phycical_meaning}
The bounds in Eqs.~\eqref{eq:main_cov_bound}--\eqref{eq:main_full_bound} hold for any passive Gaussian sensor for a given $k$ input SMSS.  We now show that these bounds are also tight and that there exist passive families of such networks for which the leading Heisenberg-scaling QFIM reaches the maximum allowed ranks stated in these bounds.

The covariance bound~\eqref{eq:main_cov_bound} is saturated when the symmetric blocks $C_i=PB_iP$ in Eq.~\eqref{eq:covariance_Gram} span the full space $\mathrm{Sym}(k,\mathbb R)$. Physically, these
directions correspond to the $k$ independent squeezed variances and the
$k(k-1)/2$ independent pairwise covariance directions inside the squeezed block. Since $\dim\mathrm{Sym}(k,\mathbb R)=k(k+1)/2$, no passive network can give more Heisenberg-scaling directions in the covariance contribution using $k$ squeezed inputs.

The first-moment bound in Eq.~\eqref{eq:main_mean_bound} is saturated when the parameter-induced displacements have \emph{all} $k$ independent parameter combinations within the squeezed subspace at Heisenberg order. This means that the vectors $\bm g_i = P(B_i\bm q + A_i\bm p)$, associated with the coefficient matrix $\mathcal F_{\rm HS}^{\bm \mu}$ in Eq.~\eqref{eq:mean_Gram_vector}, describe how the $i$th parameter shifts the mean field projected onto the squeezed modes. The bound is saturated when these vectors span $\mathbb R^k$.

For the full bound in Eq.~\eqref{eq:main_full_bound}, it is not enough to saturate the two bounds separately and is saturated only if the covariance and first-moment contributions together provide $k(k+3)/2$ independent Heisenberg-scaling directions. Once the covariance contribution already uses all $k(k+1)/2$ independent directions, the first-moment contribution must add $k$ genuinely \emph{different} independent directions from the covariance directions. Therefore, the placement of the coherent state is crucial here. If the coherent displacement lies entirely within the input squeezed subspace, then the parameter-induced mean shifts generated by the same couplings that already rotate the squeezed covariance cannot, in general, give an additional $k$ independent directions beyond those already encoded in the covariance. By contrast, placing the coherent displacement in an unsqueezed input mode provides a sufficient condition for saturating the full bound. A sufficient condition for the saturation of the full bound is derived in section~\ref{app:full_saturation_condition} of SM.


We now show that the bounds in Eqs.~\eqref{eq:main_cov_bound},~\eqref{eq:main_mean_bound} and~\eqref{eq:main_full_bound} are sharp by constructing a finite passive optical network whose generators separate the two contributions of the QFIM, achieving the Heisenberg-scaling precision in the parameter combinations. 
\begin{proposition}[Optical network for saturation of the bounds] \label{prop:saturation_main} Given $n_{\rm HS}=k(k+3)/2$ independent combinations of unknown parameters, one can build a passive optical network,
\begin{equation} 
U(\bm\phi) = 
U_0 \prod_{\ell=1}^{n_{\rm HS}} \exp\!\left(\phi_\ell G_\ell\right), 
\label{eq:sat_unitary_basis} 
\end{equation} 
where $U_0$ is a parameter-independent unitary, and $G_\ell=A_\ell+iB_\ell$ are $n_{\rm HS}$ passive generators whose phase-space representation $\tau_\ell$ is defined by the block structure of Eq.~\eqref{Eq:Generator_tau}. For $k$-SMSS into the first $k$ input ports and a coherent state into the $(k+1)$th input mode, there exists a choice of passive family of generators $ G_\ell$ such that 
\begin{equation} 
n^{\Gamma}_{\rm HS} = \frac{k(k+1)}{2}, \quad n^{\bm\mu}_{\rm HS} = k, \quad n_{\rm HS} = \frac{k(k+3)}{2}.
\label{eq:sat_full_rank_statement} \end{equation} 
Hence the covariance bound, the first-moment bound, and the bound on the full QFIM in Eqs.~\eqref{eq:main_cov_bound},~\eqref{eq:main_mean_bound} and~\eqref{eq:main_full_bound} are saturated simultaneously. Consequently, the bound in Corollary~\ref{cor:min_squeezers_full_hs} is also saturated.
\end{proposition}

One such choice is obtained by constructing the symmetric blocks \(B_\ell\) such that the first $k(k+1)/2$ symmetric blocks $B_\ell$ span all symmetric directions inside the squeezed input subspace, 
\begin{equation} B_{aa}=E_{aa}, \qquad B_{ab} = \frac{E_{ab}+E_{ba}}{\sqrt{2}}, \qquad 1\le a<b\le k , \label{eq:sat_sym_basis} \end{equation} 
and the remaining $k$ blocks couple the coherent state in $k+1$th mode to the squeezed modes,
\begin{equation} B_{k(k+1)/2+a} = E_{a,k+1}+E_{k+1,a}, \qquad a=1,\ldots,k . \label{eq:sat_mean_couplers}
\end{equation}
Here $E_{ab}$ denotes the $M\times M$ elementary matrix, $(E_{ab})_{ij}=\delta_{ia}\delta_{jb}$, with input modes ordered such that $1,\ldots,k$ modes are squeezed and $k+1$th mode carries the coherent state. Since the covariance contribution to the Heisenberg scaling determined by $C_\ell=PB_\ell P$ in Eq.~\eqref{eq:covariance_Gram}, saturating the covariance bound~\eqref{eq:main_cov_bound} requires the projected blocks $C_\ell$ to span $\mathrm{Sym}(k,\mathbb R)$. The choice in Eq.~\eqref{eq:sat_sym_basis} is precisely the Hilbert-Schmidt orthonormal basis of this space, where $E_{aa}$ gives the diagonal variance directions, while $(E_{ab}+E_{ba})/\sqrt{2}$ gives the off-diagonal covariance directions. The symmetric blocks $B_\ell$ in Eq.~\eqref{eq:sat_mean_couplers} instead couple the coherent-state mode to each squeezed mode, producing the $k$ independent Heisenberg-scaling directions that saturate the first-moment contribution bound in Eq.~\eqref{eq:main_mean_bound}. A detailed proof of Proposition~\ref {prop:saturation_main} for such a choice of passive unitary is given in Sec.~\ref{APP:Prrof_Prop3} of the SM.

In this passive linear-optical network, the 
diagonal generators $B_{aa}$  correspond to single-mode phase shifts on individual mode in the $k$ squeezed directions, whereas $B_{ab}$ 
and generators $B_{k(k+1)/2+a}$ describe two-mode couplings implemented with beam splitters and phase shifters. The resulting unitary, or any equivalent family of optical networks spanning the same independent directions, can be realized using a Reck or Clements decomposition~\cite{reck1994experimental,clements2016optimal}.

\section{Conclusion}
We have shown that Heisenberg scaling in passive multimode Gaussian metrology is governed
by a finite-dimensional geometry, namely the active squeezed input subspace.

 With $k$ single-mode squeezed inputs, the covariance contribution can support at most $k(k+1)/2$ Heisenberg-scaling directions. Meanwhile, the first-moment contribution can add at most $k$ further directions when a coherent displacement resource also scales with the total photon number. Therefore, the full QFIM has at most $k(k+3)/2$ independent Heisenberg-scaling parameter combinations. Equivalently, it sets the minimum number of squeezed inputs needed to estimate all $p$ parameters in an $M$-channel network with Heisenberg scaling.

This provides a practical resource-counting principle for large photonic sensors. Increasing the number of modes, phases, or tunable optical elements does not, in itself, increase the number of directions that achieve Heisenberg scaling. This number is determined by how the available squeezed inputs couple to the encoded parameters through the covariance and first moments.
We have also demonstrated that these bounds are sharp and can be implemented by finite passive interferometers. The resulting geometric limits therefore provide both a fundamental resource bound and a concrete benchmark for multimode Gaussian quantum sensors.

\acknowledgments
VT acknowledges partial support from the Air Force Office of Scientific Research under award number FA8655-23-17046. PF was partially supported by Istituto Nazionale di Fisica Nucleare (INFN) through the project ``QUANTUM'', by the Italian National Group of Mathematical Physics (GNFM-INdAM), and by the Italian funding within the ``Budget MUR - Dipartimenti di Eccellenza 2023--2027'' - Quantum Sensing and Modelling for One-Health (QuaSiModO).

\twocolumngrid 
\bibliography{mybib}
\clearpage
\onecolumngrid
\setcounter{section}{0}
\setcounter{equation}{0}
\setcounter{figure}{0}
\setcounter{table}{0}

\renewcommand{\thesection}{\Roman{section}}
\renewcommand{\theequation}{S\arabic{equation}}
\renewcommand{\thefigure}{S\arabic{figure}}
\renewcommand{\thetable}{S\arabic{table}}

\makeatletter
\renewcommand{\theHsection}{S\Roman{section}}
\renewcommand{\theHequation}{S\arabic{equation}}
\renewcommand{\theHfigure}{S\arabic{figure}}
\renewcommand{\theHtable}{S\arabic{table}}
\makeatother
\begin{center}
{\large\bf Supplemental Material}

\vspace{0.4cm}

{\bf Geometric bounds on multiparameter Heisenberg scaling in optical metrology with limited squeezed resources}

\vspace{0.4cm}
Atmadev Rai,$^{1,2}$ Paolo Facchi,$^{3,4}$ and Vincenzo Tamma$^{1,2,5}$

\vspace{0.2cm}

{\small
$^{1}$School of Mathematics and Physics, University of Portsmouth, Portsmouth PO1 3QL, United Kingdom\\
$^{2}$Quantum Science and Technology Hub, University of Portsmouth, Portsmouth PO1 3QL, United Kingdom\\
$^{3}$Dipartimento di Fisica, Universit\`{a} di Bari \textup{\&} Politecnico di Bari, I-70126 Bari, Italy\\
$^{4}$INFN, Sezione di Bari, I-70126 Bari, Italy\\
$^{5}$Institute of Cosmology and Gravitation, University of Portsmouth, Portsmouth PO1 3FX, United Kingdom
}
\end{center}

\vspace{0.5cm}

\section{\texorpdfstring{Asymptotic form of the Fisher matrix in Eq.~(6)}{{Asymptotic form of the Fisher matrix in Eq.(6)}}}
\label{app:HS_scaling}
In this section, we derive the asymptotic QFIM at the level of first and second moments explicitly given in Eqs.~\eqref{eq:F_cov_main} and \eqref{eq:F_mean_main}. The matrix elements of the QFIM for the pure Gaussian model are given as
\begin{align}
\mathcal F^{\Gamma}_{ij}
&=
\frac{1}{4}
\Tr\!\left[
(\partial_i\Gamma\,\Gamma^{-1})
(\partial_j\Gamma\,\Gamma^{-1})
\right],
\label{eq:App_F_cov_main}
\\
\mathcal F^{\bm \mu}_{ij}
&=
(\partial_i\bm \mu)^\mathsf{T}
\Gamma^{-1}
(\partial_j\bm \mu).
\label{eq:App_F_mean_main}
\end{align}

First, we take the input covariance in the form
\begin{equation}
\Gamma_{\rm in}
=
\frac12\begin{pmatrix}
d&0\\
0&d^{-1}
\end{pmatrix},
\qquad
d=\operatorname{diag}(\underbrace{s,\dots,s}_{k},1,\dots,1),
\qquad
s=e^{2r},
\label{Eq:app_K_in}
\end{equation}
thus the first $k$ input modes are prepared in \(p\)-squeezed
single-mode squeezed vacuum states, while the remaining modes are unsqueezed and may carry coherent displacements or vacuum.
Note that if the active squeezed modes are not the first $k$ channels but an arbitrary subset $S\subset\{1,\dots, M\}$ of size $k$, one can always write by permutation of the input so the problem is back in the form of Eq.~\eqref{Eq:app_K_in}. From
\begin{equation}
\Gamma=R_U\Gamma_{\rm in}R^\mathsf{T}_U,
\qquad
\Gamma^{-1}=R_U\Gamma_{\rm in}^{-1}R^\mathsf{T}_U,
\end{equation}
and by defining an operator $\tau_i:=R^\mathsf{T}_U\partial_iR_U$, one finds
\begin{align}
\partial_i\Gamma
&=
(\partial_iR_U)\Gamma_{\rm in}R^\mathsf{T}_U+R_U\Gamma_{\rm in}(\partial_iR_U)^\mathsf{T}
\nonumber\\
&=
R_U\tau_i\Gamma_{\rm in}R^\mathsf{T}_U+R_U\Gamma_{\rm in}\tau_i^\mathsf{T}R^\mathsf{T}_U
\nonumber\\
&=
R_U\big(\tau_i\Gamma_{\rm in}-\Gamma_{\rm in}\tau_i\big)R^\mathsf{T}_U.
\end{align}
Hence
\begin{equation}
\partial_i\Gamma\,\Gamma^{-1}
=
R_U\big(\tau_i-\Gamma_{\rm in}\tau_i\Gamma_{\rm in}^{-1}\big)R^\mathsf{T}_U,
\end{equation}
and by the cyclicity of the trace, $\mathcal F^\Gamma$ in Eq.~\eqref{eq:App_F_cov_main} reads
\begin{equation}
\mathcal F^\Gamma_{ij}
=
\frac{1}{4}\Tr\!\left[
\big(\tau_i-\Gamma_{\rm in}\tau_i\Gamma_{\rm in}^{-1}\big)
\big(\tau_j-\Gamma_{\rm in}\tau_j\Gamma_{\rm in}^{-1}\big)
\right],
\label{eq:app_FGamma_pullback}
\end{equation}
which is given by Eq.~\eqref{eq:F_cov_main} in the main text. By defining $M_i:=\tau_i-\Gamma_{\rm in}\tau_i\Gamma_{\rm in}^{-1}$, we can write 
\begin{equation}
\mathcal F^\Gamma_{ij}=\frac{1}{4}\Tr[M_i\,M_j].
\label{eq:App_FIM_Gamma_Projector1}
\end{equation}
Since $\tau_i$ symplectic generator and gives $\tau_i^\mathsf{T}=-\tau_i$ (as $R^\mathsf{T}_UR_U=\mathbb I$), $\tau_i$ takes the block form
\begin{equation}
\tau_i=
\begin{pmatrix}
A_i & -B_i\\
B_i & A_i
\end{pmatrix},
\qquad
A_i^\mathsf{T}=-A_i,
\qquad
B_i^\mathsf{T}=B_i.
\label{eq:app_tau_form}
\end{equation}
Using Eq.~\eqref{eq:app_tau_form}, one obtains
\begin{equation}
M_i=
\begin{pmatrix}
A_i-dA_id^{-1} & -B_i+dB_id\\[1mm]
B_i-d^{-1}B_id^{-1} & A_i-d^{-1}A_id
\end{pmatrix}.
\label{eq:app_Mi_blocks}
\end{equation}

Let
\begin{equation}
P=\operatorname{diag}(\underbrace{1,\dots,1}_{k},0,\dots,0),
\end{equation}
denote the projectors onto the active squeezed subspace and its orthogonal complement $Q=\mathbb I_M-P$. Then
\begin{equation}
d=sP+Q,
\qquad
d^{-1}=s^{-1}P+Q.
\label{eq:app_d_projectors}
\end{equation}

We first obtain the leading large-$s$ contribution to $\mathcal F^\Gamma$.

Using Eq.~\eqref{eq:app_d_projectors} we get
\begin{align}
dB_id
&=
(sP+Q)B_i(sP+Q)
\nonumber\\
&=
s^2PB_iP+sPB_iQ+sQB_iP+QB_iQ,
\label{eq:app_dBd}
\\[1mm]
d^{-1}B_id^{-1}
&=
(s^{-1}P+Q)B_i(s^{-1}P+Q)
\nonumber\\
&=
s^{-2}PB_iP+s^{-1}PB_iQ+s^{-1}QB_iP+QB_iQ,
\label{eq:app_dinvBdinv}
\\[1mm]
dA_id^{-1}
&=
(sP+Q)A_i(s^{-1}P+Q)
\nonumber\\
&=
PA_iP+sPA_iQ+s^{-1}QA_iP+QA_iQ,
\label{eq:app_dAdinv}
\\[1mm]
d^{-1}A_id
&=
(s^{-1}P+Q)A_i(sP+Q)
\nonumber\\
&=
PA_iP+s^{-1}PA_iQ+sQA_iP+QA_iQ.
\label{eq:app_dinvAd}
\end{align}

Therefore
\begin{align}
-B_i+dB_id
&=
(s^2-1)PB_iP+(s-1)(PB_iQ+QB_iP),
\label{eq:app_upper_right}
\\[1mm]
B_i-d^{-1}B_id^{-1}
&=
(1-s^{-2})PB_iP+(1-s^{-1})(PB_iQ+QB_iP),
\label{eq:app_lower_left}
\\[1mm]
A_i-dA_id^{-1}
&=
(1-s)PA_iQ+(1-s^{-1})QA_iP,
\label{eq:app_upper_left}
\\[1mm]
A_i-d^{-1}A_id
&=
(1-s^{-1})PA_iQ+(1-s)QA_iP.
\label{eq:app_lower_right}
\end{align}

It is clear from the above Equations that the only term carrying $O(s^2)$ ($s^2=\e^{4r}\sim N_s^2$, which leads to Heisenberg scaling) is the symmetric block
\begin{equation}
C_i:=PB_iP.
\label{eq:app_Ci_def}
\end{equation}
All other blocks are at most $O(s)$, thus $M_i$ asymptotically reads
\begin{equation}
M_i \sim
\begin{pmatrix}
-s P A_i Q & s^2 C_i\\
C_i + P B_i Q + Q B_i P& -s Q A_i P
\end{pmatrix}.
\label{eq:app_Mi_lead}
\end{equation}
Substituting Eq.~\eqref{eq:app_Mi_lead} into Eq.~\eqref{eq:App_FIM_Gamma_Projector1} gives
\begin{align}
\mathcal F^\Gamma_{ij}
&=
\frac{1}{4}\Tr\left[M_iM_j\right]
\nonumber\\
&=
\frac{1}{4}\Tr\!\left[
\begin{pmatrix}
s^2 P B_i P B_j& \cdots\\
\cdots & s^2 B_i P B_jP
\end{pmatrix}
\right]
+O(s)
\nonumber\\
&=
\frac{1}{4} s^2 \Tr\left[P B_i P B_j + B_i P B_j P \right]+O(s)
\nonumber\\
&= \frac{1}{2}s^2\Tr\left[C_iC_j\right]+O(s).
\label{eq:app_FGamma_asym_s}
\end{align}
Since \(N_s=k\sinh^2 r\), therefore for the large-squeezing limit
\(s=e^{2r}\sim 4N_s/k\), we get
\begin{equation}
\mathcal F^\Gamma_{ij}
=\frac{8N_s^2}{k^2} \Tr\left[C_iC_j\right]+O(N_s),
\label{eq:app_FGamma_final}
\end{equation}
and recover Eq.~\eqref{eq:asym_cov_main} in the main text. The leading Heisenberg scaling coefficient matrix is defined by
\begin{equation}
\mathcal F^\Gamma_{\rm HS}:=
\frac{8}{k^2} \big[\Tr\left[C_iC_j\right]\big]_{i,j=1}^{p}.
\label{eq:app_FGamma_HS_def}
\end{equation}
The matrices $C_i$ are real symmetric $k\times k$ matrices $C_i\in \mathrm{Sym}(k,\mathbb R)$ (as $B_i$ is symmetric).

\section{Proof of Proposition~\ref{prop:cov_bound_main}}
\label{App:Proposition_1}
The proof of the proposition~\ref{prop:cov_bound_main} follows directly from Eq.~\eqref{eq:app_FGamma_HS_def}.
The matrices $C_i$ are real symmetric $k\times k$ matrices
$C_i\in \mathrm{Sym}(k,\mathbb R)$.
Hence $\mathcal F^\Gamma_{\rm HS}\sim\Tr\left[C_iC_j\right]$ is proportional to the Gram matrix of the family $\{C_i\}$ with respect to the Hilbert--Schmidt inner product
\begin{equation}
\langle X,Y\rangle:=\Tr\left[X\,Y\right].
\end{equation}
Therefore, by using the property of a Gram matrix, 
\begin{equation}
\operatorname{rank}\!\big(\mathcal F^\Gamma_{\rm HS}\big)
=
\dim\operatorname{span}\{C_i\}
\le
\dim\mathrm{Sym}(k,\mathbb R).
\label{eq:app_rank_cov_step}
\end{equation}
Since
\begin{equation}
\dim\mathrm{Sym}(k,\mathbb R)
=
k+\frac{k(k-1)}{2}
=
\frac{k(k+1)}{2},
\label{eq:app_dim_sym}
\end{equation}
and, as the rank of the QFIM cannot exceed the total number of parameters in the channel,  $\operatorname{rank}(\mathcal F^\Gamma_{\rm HS})\le p$, we conclude that 
\begin{equation}
n_{\mathrm{HS}}^{\Gamma} = \operatorname{rank}\!\big(\mathcal F^\Gamma_{\rm HS}\big)
\le
\min\!\left\{p,\frac{k(k+1)}{2}\right\},
\end{equation}
which proves Proposition \ref{prop:cov_bound_main}.

\section{\texorpdfstring{Asymptotic form of $\mathcal{F}^{\bm \mu}$ in Eq.~(8) and proof of Proposition~\ref{prop:mean_bound_main}}{Asymptotic form of F in Eq.(8) and proof of Proposition}}
\label{app:HS_scaling_mean}

In this section, we analyze the first-moment contribution of the Fisher matrix and show that it scales at leading order as $N_cN_s$ and hence achieves the Heisenberg scaling for $N_{s,c}=O(N)$. Later in this section, we also give the proof of Proposition~\ref{prop:mean_bound_main} in the main text.

The input mean vector is written as
\begin{equation}
\bm\mu_{\rm in}=\sqrt{2N_c}\binom{\bm q}{\bm p},
\qquad
\|\bm q\|^2+\|\bm p\|^2=1,
\end{equation}
where $N_c$ denotes the total coherent-displacement resource, while the normalized vectors $\bm q=(q_1,\dots,q_M)$ and $\bm p=(p_1,\dots,p_M)$ specify how the displacement is distributed across the $M$ input modes. Here, for generalization, no restriction is imposed on the number or locations of the displaced inputs.
From 
\begin{equation}
\bm{\mu}=R_U\,\bm{\mu_{\rm in}},
\qquad
\partial_i\bm{\mu}=(\partial_iR_U)\,\bm{\mu_{\rm in}}=R_U\,\tau_i\, \bm{\mu_{\rm in}},
\end{equation}
we can write the Fisher matrix $\mathcal F^{\bm\mu}$ in~\eqref{eq:F_mean_main} as
\begin{equation}
\mathcal F^{\bm{\mu}}_{ij}=(\partial_i\bm \mu)^\mathsf{T}
\Gamma^{-1}
(\partial_j\bm \mu)
=
\bm{\mu}_{\rm in}^\mathsf{T}\, \tau_i^\mathsf{T}\, \Gamma_{\rm in}^{-1}\,\tau_j \,\bm{\mu}_{\rm in}.
\label{eq:Fmu_exact}
\end{equation}
Using the property of $\tau_i$ from Eq.~\eqref{eq:app_tau_form}, we obtain
\begin{equation}
\tau_i\binom{\bm q}{\bm p}=\begin{pmatrix}
    A_i&-B_i\\B_i&A_i
\end{pmatrix}\binom{\bm q}{\bm p}
=
\binom{\xi_i}{\eta_i},
\qquad
\xi_i=A_i\bm q-B_i\bm p,
\qquad
\eta_i=B_i\bm q+A_i\bm p.
\end{equation}
Hence
\begin{equation}
\mathcal F^{\bm{\mu}}_{ij}
=
4N_c
\left(
\xi_i^\mathsf{T} d^{-1}\xi_j+\eta_i^\mathsf{T} d\,\eta_j
\right).
\label{eq:Fmu_exact_xi_eta}
\end{equation}

Using Eq.~\eqref{eq:app_d_projectors}, we get
\begin{align}
\mathcal F^{\bm \mu}_{ij}
&=
4N_c
\left[
\xi_i^\mathsf{T}(s^{-1}P+Q)\xi_j
+
\eta_i^\mathsf{T}(sP+Q)\eta_j
\right]
\nonumber\\
&=
4N_c
\left[
s^{-1}(P\xi_i)^\mathsf{T}(P\xi_j)
+
(Q\xi_i)^\mathsf{T}(Q\xi_j)
+
s(P\eta_i)^\mathsf{T}(P\eta_j)
+
(Q\eta_i)^\mathsf{T}(Q\eta_j)
\right].
\label{eq:app_Fmu_split}
\end{align}
From the above equation, it is clear that the only term that can contribute to Heisenberg scaling ($s N_c= N_c\e^{2r}\sim 4 N_cN_s$) is the third term in the equation $s(P\eta_i)^\mathsf{T}(P\eta_j)$, given that $N_{s,c}=O(N)$. Let us define
\begin{equation}
\bm g_i:=P\eta_i=P(B_i\bm q+A_i\bm p)\in\mathbb R^k.
\label{eq:app_gi_def}
\end{equation}
Then the Fisher information matrix $\mathcal F^{\bm \mu}$ reads 
\begin{equation}
\mathcal F^{\bm \mu}_{ij}=4sN_c\, \bm{g}_{i}^\mathsf{T} \bm g_j + O(N_c + s) =
\frac{16\,N_sN_c}{k}\,\left( \bm g_i^{\mathsf T} \bm g_j\right) + O(N_c + N_s),
\label{Neweq:app_Fmu_lead}
\end{equation}
which is Eq.~\eqref{eq:asym_mean_main} in the main text. To contribute to an order $N^2$, the displacement resource $N_c$ must scale linearly with the squeezing resource $N_s$, say $N_s=\beta N,\; N_c=(1-\beta)N$, where $0\le \beta \le 1$. Eq.~\eqref{Neweq:app_Fmu_lead} asymptotically becomes 
\begin{equation}
\mathcal F^{\bm \mu}_{ij}=\frac{16}{k}\beta(1-\beta) N^2\, \bm g_i^{\mathsf T} \bm g_j + O(N).
\label{eq:App_FIM_mean}
\end{equation}
This defines the leading first-moment coefficient matrix in Eq.~\eqref{eq:asym_mean_main}
\begin{equation}
\mathcal F^{\bm \mu}_{\rm HS}:=
\frac{16}{k} [\bm g_i^{\mathsf T} \bm g_j]_{i,j=1}^{p}.
\label{eq:app_Fmu_HS_def}
\end{equation}

\subsection{Proof of Proposition~\ref{prop:mean_bound_main}}
We see that the above coefficient matrix $\mathcal{F}_{\rm HS}^{\bm \mu}\sim [\bm g_i^{\mathsf T} \bm g_j]$ is again proportional to a Gram matrix, now of the family $\{\bm g_i\}\subset \mathbb R^k$. Hence
\begin{equation}
\operatorname{rank}\!\big(\mathcal F^{\bm \mu}_{\rm HS}\big)
=
\dim\operatorname{span}\{\bm g_i\}
\le
k.
\end{equation}
By including the trivial bound $\operatorname{rank}(\mathcal F^{\bm \mu}_{\rm HS})\le p$, we obtain
\begin{equation}
\operatorname{rank}\!\big(\mathcal F^{\bm \mu}_{\rm HS}\big)=n_{\mathrm{HS}}^{\bm{\mu}}
\le
\min\{p,k\},
\end{equation}
which proves Proposition \ref{prop:mean_bound_main}.


\section{Proof of Corollary~\ref{cor:full_bound_main} and~\ref{cor:min_squeezers_full_hs}}
\label{App:proof_of_Corollary_1}
Assume that both the covariance and first-moment contributions are present at
Heisenberg order. Combining Eqs.~\eqref{eq:app_FGamma_final} and
\eqref{eq:App_FIM_mean}, the full Fisher matrix takes the asymptotic form
\begin{equation}
\mathcal F=\mathcal F^{\Gamma}+\mathcal F^{\bm \mu}
\sim
N^2\left(\beta^2\,\mathcal F_{\rm HS}^\Gamma
+\beta(1-\beta)\mathcal{F}_{\rm HS}^{\bm \mu}\right)=N^2\mathcal F_{\rm HS},
\end{equation}
with
\begin{equation}
\mathcal F_{{\rm HS},\,ij}
=
\frac{8}{k^2}\beta^2\Tr\left[C_iC_j\right]+\frac{16}{k}\beta (1-\beta)\,\bm g_i^{\mathsf T} \bm g_j.
\label{eq:app_Ffull_HS}
\end{equation}
The precise positive prefactors are not relevant for the rank, but they show explicitly that the leading coefficient matrix is a Gram matrix on the direct-sum space $\mathrm{Sym}(k,\mathbb R)\oplus \mathbb R^k$,
since Eq.~\eqref{eq:app_Ffull_HS} can be written as the inner product of the vectors
\begin{equation}
\mathcal X_i:=(\frac{\sqrt{8}\beta}{k}C_i,
\frac{4}{\sqrt{k}}\sqrt{\beta(1-\beta)}\,\bm g_i).
\end{equation}
Therefore
\begin{equation}
\operatorname{rank}\!\big(\mathcal F_{\rm HS}\big)
\le
\dim\mathrm{Sym}(k,\mathbb R)+\dim\mathbb R^k
=
\frac{k(k+1)}{2}+k
=
\frac{k(k+3)}{2}.
\end{equation}
Since $\operatorname{rank}(\mathcal F_{\rm HS})\le p$ (as the rank of QFIM cannot exceed the number of parameters $p$ in the given unitary), we obtain
\begin{equation}
\operatorname{rank}\!\big(\mathcal F_{\rm HS}\big)=n_{\mathrm{HS}}
\le
\min\!\left\{p,\frac{k(k+3)}{2}\right\},
\end{equation}
which proves Corollary \ref{cor:full_bound_main}.

To estimate all p parameters with Heisenberg scaling, $\mathcal{F}_{\rm HS}$ needs to be of full rank, $n_{\rm HS}=\operatorname{rank}\!\big(\mathcal F_{\rm HS}\big)=p$. However, from Corollary~\ref{cor:full_bound_main},
\begin{equation}
n_{\rm HS}\le
\min\!\left\{p,\frac{k(k+3)}{2}\right\}.
\end{equation}
Hence $n_{\rm HS}=p$ is possible only if
\begin{equation}
\frac{k(k+3)}{2}\geq p.
\end{equation}
Solving the quadratic inequality $k^2+3k-2p\ge0$ gives
\begin{equation}
k_{\pm}=\frac{-3\pm\sqrt{9+8p}}{2}.
\end{equation}
Since $k$ is a non-negative integer, the relevant condition is
\begin{equation}
k\ge
\left\lceil k_+ \right\rceil
=
\left\lceil
\frac{\sqrt{8p+9}-3}{2}
\right\rceil ,
\end{equation}
which proves Corollary~\ref{cor:min_squeezers_full_hs}.

\subsection{For unequal squeezed inputs}
\label{Appendix_D1}
In this section, we show that the same conclusion holds when the $k$-SMSS are not equally squeezed, provided they scale with the same asymptotic resource. Let
\begin{equation}
N_a=\sinh^2 r_a \sim \rho_a N_s,
\qquad
\rho_a>0, \qquad \sum_{a}\rho_a=1.
\end{equation}
In this case, the leading Heisenberg-order matrices are no longer isotropic on the active subspace. Then the leading covariance matrix becomes a weighted Gram matrix on $\mathrm{Sym}(k,\mathbb R)$,
\begin{equation}
\mathcal F^{\Gamma}_{{\rm HS},ij}
=
8 
\sum_{a,b\in A}
\rho_a\rho_b\,
(C_i)_{ab}(C_j)_{ab},
\label{eq:weighted_cov_main}
\end{equation}
and, when the displacement resource satisfies $N_s=\beta N$, $N_c=(1-\beta)N$, the first-moment contribution becomes a weighted Gram matrix on the active vector space,
\begin{equation}
\mathcal F^{\bm\mu}_{{\rm HS},ij}
=
16
\sum_{a\in A}
\rho_a\,
(\bm g_i)_a(\bm g_j)_a.
\label{eq:weighted_mean_main}
\end{equation}
The positive constants $\rho_a$ only change the metric within these spaces. Thus the covariance contribution still lives in $\mathrm{Sym}(k,\mathbb R)$, whose dimension is $k(k+1)/2$, while the first-moment contribution still lives in $\mathbb R^k$, whose dimension is $k$.
This shows that unequal squeezing merely weights different directions inside the same active subspace. As long as all $k$ SMSS scale linearly with the common resource $N$, the number of independent Heisenberg-scaling directions is fixed by the dimension of that subspace.

\section{Coherent displacement and saturation of the full Heisenberg-scaling rank}
\label{app:full_saturation_condition}
In this section, we clarify the role of input coherent displacement in saturating the full bound in Corollary~\ref{cor:full_bound_main}.  The full Heisenberg-scaling QFIM is given by the Gram matrix of the combined Eqs.~\eqref{eq:app_Ci_def} and~\eqref{eq:app_gi_def}
\begin{equation}
\Lambda_i=(C_i,\bm g_i)
\in
\mathrm{Sym}(k,\mathbb R)\oplus\mathbb R^k .
\label{eq:app_full_lifted_objects}
\end{equation}
Once the covariance contribution bound is saturated, i.e., the matrices \(C_i=P B_iP\) span \(\mathrm{Sym}(k,\mathbb R)\), which has the dimension $k(k+1)/2$, the only way to saturate the full bound
\begin{equation}
n_{\rm HS}=\frac{k(k+1)}2+k
=
\frac{k(k+3)}2,
\end{equation}
is for the first-moment contribution $(n_{\rm HS}^{\bm \mu})$ to add $k$ further independent directions other than $n_{\rm HS}^{\Gamma}$. Equivalently, if $T_\Gamma:\tau_i\mapsto C_i$ and $T_\mu:\tau_i\mapsto \bm g_i$, the rank of the combined vector obeys
\begin{equation}
\dim\operatorname{span}\{\Lambda_i\}
=
\dim\operatorname{span}\{C_i\}
+
\dim T_\mu(\ker T_\Gamma).
\label{eq:app_rank_identity_kernel}
\end{equation}
The first term counts the independent directions of covariance contribution, while the quantity \(\dim T_\mu(\ker T_\Gamma)\) counts the number of independent first-moment directions that can be generated by parameter variations which do not contribute to the covariance block $n_{\rm HS}^{\Gamma}$. Therefore, after the covariance contribution has reached its maximum dimension, i.e., $\dim\operatorname{span}\{C_i\}=k(k+1)/2$, the saturation of the full-rank bound $n_{\rm HS}$ requires
\begin{equation}
\dim T_\mu(\ker T_\Gamma)=k.
\label{eq:app_full_saturation_condition}
\end{equation}
We now show why the placement of the coherent state becomes relevant for the saturation of this full-rank bound. We consider generator directions whose active covariance block vanishes, i.e., if generators \(\tau_i\in\ker T_\Gamma\), then $P B_iP=0$. By decomposing the input displacement using the projection operators $P$ and $Q$ defining the active squeezed subspace and the unsqueezed subspace, respectively, as
\begin{equation}
\bm q=P\bm q+Q\bm q,
\qquad
\bm p=P\bm p + Q\bm p.
\end{equation}
As \(P B_iP=0\), Eq.~\eqref{eq:app_gi_def} gives
\begin{equation}
\bm g_i=P B_iQ\,\bm q+P A_iP\,\bm p+P A_iQ\,\bm p.
\label{eq:app_g_decomposition_kernel}
\end{equation}
If the placement of the coherent state is confined entirely to the $k$ input squeezed subspace, then
\begin{equation}
Q\,\bm q=0,\qquad Q\,\bm p=0,
\end{equation}
and therefore
\begin{equation}
\bm g_i=P A_iP\,\bm p .
\label{eq:app_g_inside_P}
\end{equation}
Since $P A_iP$ is antisymmetric (as $A_i$ is antisymmetric), every vector $P A_iP\,\bm p$ is orthogonal to $P\,\bm p$. These vectors can span at most a $(k-1)$-dimensional subspace giving $\dim T_\mu(\ker T_\Gamma)\le k-1$, when $P\,\bm p\neq0$, and it is zero if $P\,\bm p=0$. Therefore, after the covariance contribution has already saturated, a coherent displacement confined only to the input squeezed subspace cannot in general give the full additional $k$ directions in Eq.~\eqref{eq:app_full_saturation_condition}.

By contrast, placing the coherent displacement in an unsqueezed input port allows the missing $k$ directions to be generated. For example, if the coherent displacement is placed in any unsqueezed input port, then
\begin{equation}
P\bm q=0,
\qquad
P\bm p=0,
\end{equation}
and the vector $\bm g_i$ in Eq.~\eqref{eq:app_gi_def} reduces to $\bm g_i=P B_i Q\,\bm q+P A_i Q\,\bm p$. The sufficient condition for obtaining the missing \(k\) directions is therefore
\begin{equation}
\dim\operatorname{span}
\left\{
P B_i Q\,\bm q+P A_i Q\,\bm p\right\}_{i}=k.
\label{eq:app_mixed_sufficient_condition}
\end{equation}
When Eq.~\eqref{eq:app_mixed_sufficient_condition} holds, the first-moment
Heisenberg-scaling directions span the full space \(\mathbb R^k\), independently of the
covariance directions.
 Combining these ranks with any covariance-saturating set of \(P B_iP\) gives
\begin{equation}
\operatorname{span}\{\Lambda_i\}
=
\mathrm{Sym}(k,\mathbb R)\oplus\mathbb R^k .
\end{equation}
Hence, when $p\ge k(k+3)/2$,
\begin{equation}
n_{\rm HS}
=
\frac{k(k+1)}2+k
=
\frac{k(k+3)}2 .
\end{equation}
In summary, the covariance contribution bound does not require any input coherent displacement. The bound resulting from the signal mean contribution $n_{\rm HS}^{\bm \mu}$ considered alone only requires the vectors $\bm g_i$ to span the active squeezed subspace; for this, one does not necessarily need the input displacement to be outside the active squeezed subspace. However, once the bound obtained from the covariance contribution $n_{\rm HS}^{\Gamma}$ is already saturated, achieving the full bound $n_{\rm HS}=k(k+3)/2$ requires the mean contribution $n_{\rm HS}^{\bm \mu}$ to add $k$ independent directions from $\ker T_\Gamma$. A coherent displacement in any one of the $M-k$ unsqueezed input ports, together with generators satisfying Eq.~\eqref{eq:app_mixed_sufficient_condition}, provides a sufficient condition for the saturation of the full rank bound.

\section{Proof of Proposition~\ref{prop:saturation_main}}
\label{APP:Prrof_Prop3}
Let $P$ project onto the first $k$ input modes prepared in SMSSs. We use one additional unsqueezed mode (assuming $M\ge k+1$), and inject into it a coherent state of real amplitude. In the phase space convention $\bm\mu_{\rm in} = \sqrt{2N_c}\,(\bm q,\bm p)^{\mathsf T}$, this corresponds to  $\bm q=\bm e_{k+1}$ and $\bm p=0$, where $\bm e_i$ denotes the unit vector selecting mode $i$. The coherent state is therefore initially outside the squeezed subspace. It becomes Heisenberg-sensitive only when a passive generator mixes it into a squeezed quadrature. 
At leading order, the covariance and first-moment contributions depend on $C_\ell=PB_\ell P$, and $\bm g_\ell=P(B_\ell\bm q+A_\ell\bm p)$ associated with the Gram matrices in Eqs.~\eqref{eq:covariance_Gram} and~\eqref{eq:mean_Gram_vector}, respectively. 
Since $\bm p=0$, the antisymmetric block $A_\ell$ does not enter the leading first-moment vector and gives $\bm g_\ell=PB_\ell\bm e_{k+1}$. For the present construction, we choose $A_\ell=0$, so that the saturation of the bound is determined entirely by the symmetric blocks $B_\ell$. 

The first $k(k+1)/2$ generators in Eq.~\eqref{eq:sat_sym_basis} act entirely inside the squeezed subspace and form an orthonormal basis of \(\mathrm{Sym}(k,\mathbb R)\), so that $\operatorname{Tr}(B_\alpha B_\beta)=\delta_{\alpha\beta}$, for $\alpha,\beta=1,\ldots,k(k+1)/2$.
The projected input squeezed subspace on these symmetric blocks gives, 
\begin{equation}
    C_\alpha=PB_\alpha P=B_\alpha,
    \quad
    \bm g_\alpha=PB_\alpha \bm e_{k+1}=0,
    \label{eq:sat_cov_data}
\end{equation}
where $\alpha=1,\ldots,k(k+1)/2$. Thus these generators feed only the covariance Gram matrix in Eq.~\eqref{eq:covariance_Gram}, giving $\operatorname{rank}\mathcal F^{\Gamma}_{\rm HS} =k(k+1)/2$.

The generators in Eq.~\eqref{eq:sat_mean_couplers} have no block fully contained entirely in the squeezed subspace, hence $C_{k(k+1)/2+a} = PB_{k(k+1)/2+a}P = 0$.
Their action on the coherent displacement is instead 
\begin{equation} \bm g_{k(k+1)/2+a} = PB_{k(k+1)/2+a}\bm e_{k+1} = \bm e_a, \label{eq:sat_mean_data} 
\end{equation}
for $a=1,\ldots,k$. Each coupler mixes the coherent amplitude of the $k+1$th mode into one squeezed mode. Since the vectors $e_a$ span the squeezed subspace, these directions feed only the first-moment Gram matrix in Eq.~\eqref{eq:mean_Gram_vector} and give $\operatorname{rank}\mathcal F^{\bm\mu}_{\rm HS} = k$. 

The two contributions are separated at leading order such that the first set has nonzero $C_\ell$ and zero $\bm g_\ell$ in Eq.~\eqref{eq:sat_cov_data}, while the second set has zero $C_\ell$ and nonzero $\bm g_\ell$ in Eq.~\eqref{eq:sat_mean_data}. Hence the Heisenberg-scaling coefficient matrix in Eq.~\eqref{eq:totalFIM} is block diagonal up to positive prefactors, 
\begin{equation}
\mathcal F_{\rm HS}
=
\begin{pmatrix}
c_{\Gamma} I_{k(k+1)/2} & 0\\
0 & c_{\bm\mu} I_k
\end{pmatrix},
\qquad
c_{\Gamma}>0,
\qquad
c_{\bm\mu}>0,
\label{eq:sat_block_identity}
\end{equation}
and therefore has $\operatorname{rank}\mathcal F_{\rm HS} = \frac{k(k+1)}{2}+k = \frac{k(k+3)}{2}$. This reaches the maximum allowed by the bounds and proves their sharpness.

\end{document}